\documentclass[amssymb,useAMS,prd,aps,amsmath,superscriptaddress,nofootinbib,twocolumn]{revtex4}
\usepackage{color}
\usepackage{pslatex}
\usepackage{pdfpages}
\usepackage{graphicx}
\usepackage{psfrag}
\usepackage{pdftricks}
\usepackage{multirow}
\usepackage{graphicx}

\begin{document}
\def\mlsp{m_{\chi_1^0}}
\title{Astrophysical limits on light NMSSM neutralinos}

\author{Daniel Albornoz V\'asquez}
\affiliation{LAPTH, U. de Savoie, CNRS,  BP 110,
  74941 Annecy-Le-Vieux, France.}

\author{Genevi\`eve B\'elanger}
\affiliation{LAPTH, U. de Savoie, CNRS,  BP 110,
  74941 Annecy-Le-Vieux, France.}

\author{C\'eline B\oe hm}
\affiliation{LAPTH, U. de Savoie, CNRS,  BP 110,
  74941 Annecy-Le-Vieux, France.}
\affiliation{IPPP, Ogden centre, Durham University, UK}


\begin{abstract}
It was recently shown  that light  LSP neutralinos could be found in the framework of the NMSSM.
These candidates would escape known Particle Physics constraints even though they are relatively light. 
We now investigate the astrophysical limits which 
can be set on these particles. 
We show, in particular, that the Fermi observation of dwarf spheroidal galaxies enable to constrain the parameter space associated with these candidates. 
Combined with the  XENON100 experimental limits, our results illustrate the 
complementarity between direct and indirect searches for dark matter. 
Yet, our findings also  
suggest that probing light neutralinos in the NMSSM scenario will be very difficult 
because the sensitivity of both dark matter direct and indirect detection experiments 
would have to be improved by at least six order of magnitude compared to present values in order 
to explore the entire parameter space. 
Finally, we show that the parameter space compatible with the CoGeNT signal (albeit disfavored by the XENON100 limit) is not excluded by gamma nor 
radio observations. 
\end{abstract}

\maketitle

\section{Introduction}

After several decades devoted to dark matter searches in underground, collider and spatial experiments, 
we are finally reaching an important cross road. Now that LHC is running, the Fermi and Planck experiments are delivering data and 
direct detection experiments have reached the level of sensitivity required to probe dark matter particles 
in the GeV-TeV range, we expect important developments in the dark matter field. 
In particular, it is likely that we obtain enough information in the next few years 
to determine  whether supersymmetry  manifests itself at the weak scale or not and, hence, whether neutralinos 
can constitute the dark matter. 

Until then, elucidating the nature of dark matter remains challenging. 
While ongoing experimental efforts try to close down the possible dark matter mass range, 
several dark matter direct detection experiments have announced events or signals which could point towards 
the existence of relatively light particles~\cite{Aalseth:2010vx,Bernabei:2010mq,Ahmed:2009zw}. 
Although these seem also compatible with background expectations, 
these claims have revived the theoretical interest for candidates in the GeV-10 GeV mass range and encouraged experiments to investigate the low energy range despite the lack of sensitivity at small recoil energies. 

From a theoretical point of view, motivating candidates in this mass range is not an easy task. For example, 
the lowest neutralino mass in the Minimal Supersymmetric Standard Model (MSSM) 
(with common slepton and squark masses and unification of the gluino and wino masses) now appears to be above 15 GeV~\cite{Vasquez:2010ru,Albornoz:2011}. 
Relaxing the universality condition,  scenarios where neutralinos could be 
below this value were found~\cite{Fornengo:2010mk,Scopel:2011qt,Calibbi:2011ug,Gogoladze:2010fu}, nevertheless 
the low mass range appears to be both statistically unlikely and challenged by direct detection and Higgs searches at colliders \footnote{Neutralinos lighter than 1 GeV have been shown to satisfy collider, astrophysical and cosmological constraints ~\cite{Dreiner:2009ic} but we do not consider these cases.}.
In view of these results,  other candidates were investigated, in particular those in extensions of the MSSM such as 
the neutralino in the Next-to-Minimal Supersymmetric Standard Model (NMSSM) ~\cite{Draper:2010ew,Gunion:2010dy,Cao:2011re,Kappl:2010qx,Kang:2010mh,Das:2010ww,Delahaye:2011cu}, in the MSSM with an extended Higgs sector ~\cite{Bae:2010hr}
or the right-handed sneutrino in supersymmetric extensions~\cite{Belanger:2010cd,MarchRussell:2009aq}.  

In a previous study, we demonstrated that NMSSM neutralinos could be very light~\cite{Vasquez:2010ru}. 
In particular, we found that many points with a LSP mass between 1 and 15 GeV were actually compatible with known particle physics constraints and possibly also with direct detection exclusion bounds. 
Here, we investigate whether these scenarios respect  the most recent astrophysical limits since 
light (1-15 GeV) particles could overproduce the radio emission in the Milky Way (MW) and in galaxy clusters \cite{Boehm:2002yz,Boehm:2010kg}, gamma rays in dwarf spheroidal (dSph) galaxies \cite{Strigari:2006rd} 
and antiprotons in the Milky Way \cite{Lavalle:2010yw,Delahaye:2011cu}. 

In Section II, we recall the parameter space associated with light NMSSM neutralinos (before applying the astrophysical and direct detection limits). We show, in particular, that some points have a very large annihilation cross section at small dark matter velocity despite a non negligible relic density. In Section III, we show how astrophysical and experimental limits cut into the parameter space, thus demonstrating the complementarity between direct and indirect detection searches. In Section IV, we estimate the radio flux expected for benchmark points,  selected so as to evade the Fermi and direct detection limits. We conclude in Section V.

\section{Neutralino annihilation in the NMSSM}
In this section, we delineate the parameter space associated with light NMSSM neutralinos and compute the dominant 
branching ratios. 
For this purpose, we consider thermal candidates and require that their energy density today 
is either equal to (or smaller than) the observed dark matter abundance, that is  
$\Omega_{\rm WMAP} h^2 > \Omega_\chi h^2 > 10\% \Omega_{\rm WMAP} h^2$ with 
$\Omega_{\rm WMAP} h^2=0.1131\pm 0.0034$~\cite{Komatsu:2008hk}. 
This basically constrains the 
total annihilation cross section of neutralinos in the primordial Universe and rules out part of the parameter space.
We also impose constraints from new particle searches at colliders, from B-physics observables and  from the muon anomalous magnetic
moment as detailed in~\cite{Vasquez:2010ru}.

Once we obtained the configurations which satisfy all these constraints, we can predict the energy spectrum of photons in the galaxy and dSph as well as the flux of cosmic rays in the MW. 
The comparison of the spin-independent cross section  with the limits from CDMS~\cite{Ahmed:2009zw} and XENON100~\cite{Aprile:2011hi} will enable us to set additional constraints. In view of the tension at low dark matter mass between the results from the XENON, CDMS, CoGeNT \cite{Aalseth:2010vx} and DAMA/LIBRA \cite{Bernabei:2010mq} experiments, we will also investigate whether some scenarios can fall into the low mass and large cross section region or not.

To perform this analysis, we have used the same Markov Chain Monte-Carlo code presented in \cite{Vasquez:2010ru} based on the NMSSMTools package~\cite{Ellwanger:2005dv} embedded in micrOMEGAs~\cite{Belanger:2005kh}. Dark matter observables were computed 
with micrOMEGAs~\cite{Belanger:2008sj,Belanger:2010gh}.

\subsection{The model}

\begin{figure}
	\centering
		\includegraphics[width=0.47\textwidth]{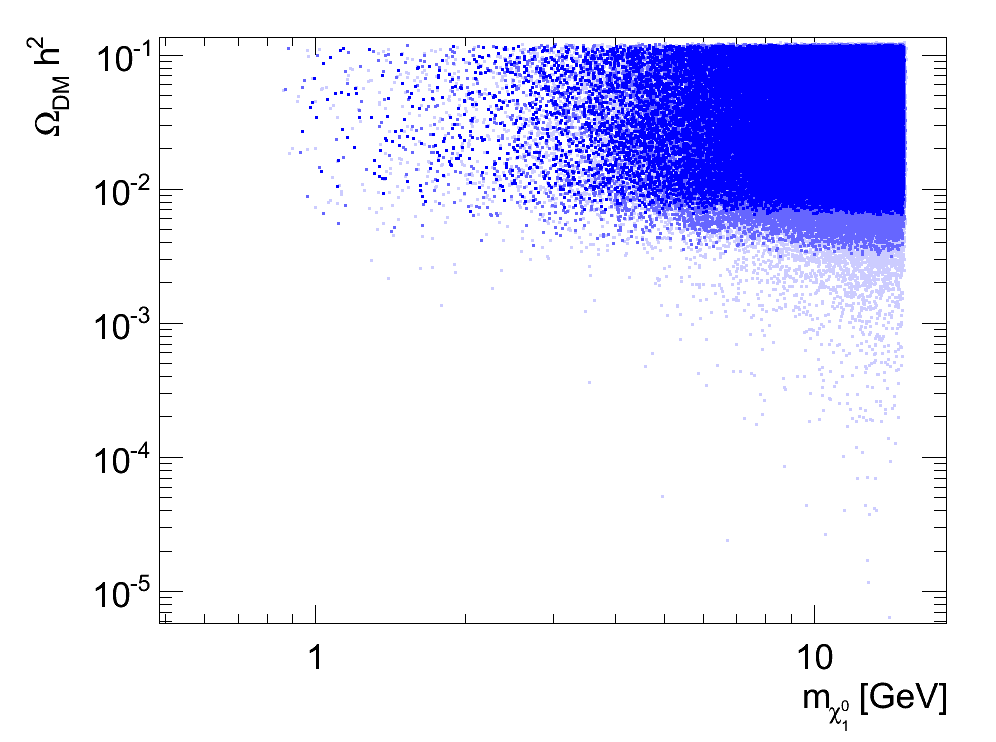}
	\caption{Relic density of light NMSSM neutralinos, the darker the dot the larger the likelihood. }
\label{fig:sigmavsmchiEU}
\end{figure}

The Next-to-Minimal Supersymmetric Standard Model
(NMSSM) is a simple extension of the MSSM that
contains an additional gauge singlet superfield. The 
VeV of this singlet induces an effective $\mu$ term that is 
 naturally of the order of the electroweak scale, thus providing a solution to the naturalness problem~\cite{Ellwanger:2009dp}.
The model contains one additional neutralino state, the singlino, as well as three scalar ($H_1,H_2,H_3$) and two pseudoscalar ($A_1,A_2$) Higgs
bosons. 
An important feature of the model is that 
the singlet fields can be very light and  escape the LEP
bounds. This is because these fields mostly decouple from
the SM fields~\cite{Ellwanger:2009dp}. 
This opens up the possibility for new annihilation
mechanisms for light neutralinos, in particular
through the  exchange of light 
Higgs singlets as well as into light Higgs singlets~\cite{Belanger:2005kh}.
The model that we consider has input parameters which are defined at the weak scale. 
The free parameters are taken to be the gaugino
masses $M_1,M_2=M_3/3$, the Higgs sector parameters $\mu,\tan\beta$, $\lambda,\kappa,A_\lambda,A_\kappa$,
 a common mass for the sleptons $m_{\tilde l}$ and the squarks $m_{\tilde q}$ as well as  only one non-zero trilinear
coupling, $A_t$, for more details see ~\cite{Vasquez:2010ru}. We only consider scenarios with a neutralino LSP lighter than  15~GeV.

\subsection{Branching ratios in the Early Universe}

We start by computing the relic density for each candidate. 
The main assumption in these calculations is the conventional freeze-out mechanism. 
As shown, in Fig.~\ref{fig:sigmavsmchiEU}, 
we have found many points with a relic density in the WMAP range.
Hence, although the candidates that we are interested in are light, many have an acceptable relic density. 
In what follows, we do also include the points with a smaller relic density even though they can only partially contribute to the dark matter.

\begin{figure}[h]
	\centering
		\includegraphics[width=0.5\textwidth]{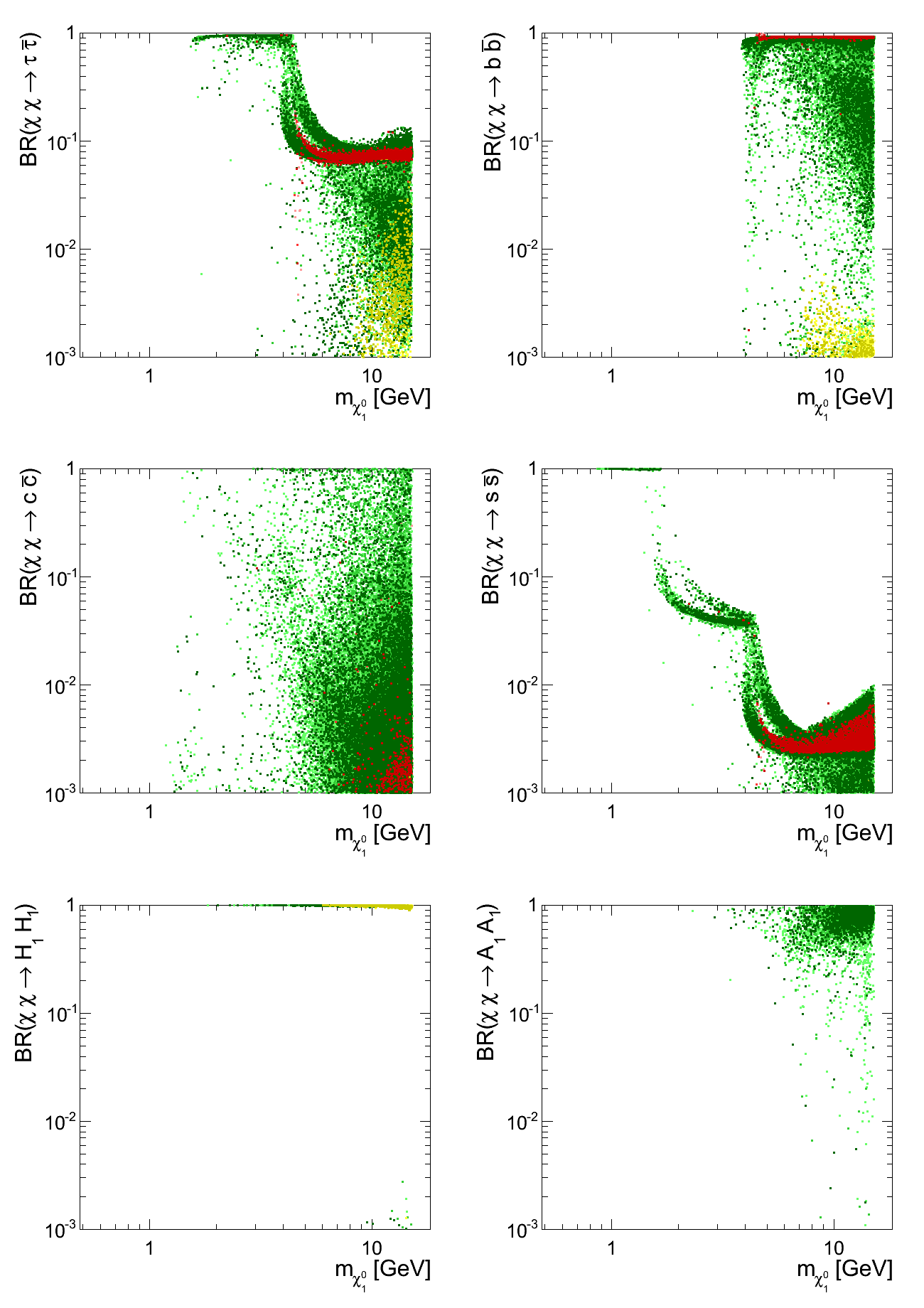}
	\caption{Branching ratios associated with neutralino pair annihilations in the Early Universe.
	The color code is associated with the constraints from dSph (red) and direct detection (yellow), see section III.}
\label{fig:BR_EU}
\end{figure}

We can now determine the Branching Ratios (BR) associated with the different final states. The results 
are displayed in Fig.~\ref{fig:BR_EU}. The dominant annihilation channel is either
into Higgs pairs, $H_1H_1$ and $A_1A_1$, or into Fermion pairs.
We did not find any configurations with $H_1A_1$ in the final state as it would
require that both the scalar and pseudoscalar be very light.  
For Fermionic modes, the dominant channel is determined by the 
heaviest kinematically accessible Fermion. When the mass of the neutralino is smaller than 1.7 GeV, 
the only possible final states are into light quarks  $s \bar{s}$ and $c \bar{c}$. 
 If the neutralino mass is larger than 1.7 GeV but smaller than 4.2 GeV 
($m_b$), the dominant channel is $\tau \bar{\tau}$ at 90-100 $\%$. 
Above the $b$ mass, the dominant Fermionic final state is usually  $b \bar{b}$.
However, the associated branching ratio spans from below 1$\%$ to 100$\%$ as the Higgs mode can also contribute significantly.
Hence, for neutralino masses above the 
b-quark mass, one expects also  annihilations into  $H_1H_1,A_1 A_1$ 
as well as some contribution from $\tau \bar{\tau}$ and $c \bar{c}$.

\subsection{Branching ratios in  Milky Way and Dwarf Spheroidal galaxies}

We can now compute the annihilation cross section and branching ratios in the MW and  dwarf spheroidal galaxies. 
In effect, this is equivalent to studying the impact of the dark matter velocity. 
To take into account the fact that neutralinos may not be responsible for all of the dark matter in the universe and therefore that the
halo may not be totally composed of neutralinos, we introduce the
parameter 
\begin{eqnarray}
\xi&=& 1 \;\;\; {\rm if}  \;\;\; \Omega_\chi h^2 \geq \Omega_{\rm WMAP} h^2\nonumber\\
&=& \Omega_\chi h^2/\Omega_{\rm WMAP} h^2  \;\;\;\;\;{\rm otherwise}
\end{eqnarray}
where $\Omega h^2_{\rm WMAP}$ corresponds to the lower value of the WMAP measurement.
In what follows we will always rescale the local neutralino density by the 
factor, $\rho_{\chi_0}=\xi \rho_{\rm DM}$.  

\begin{figure}[h]
	\centering
\includegraphics[width=0.5\textwidth]{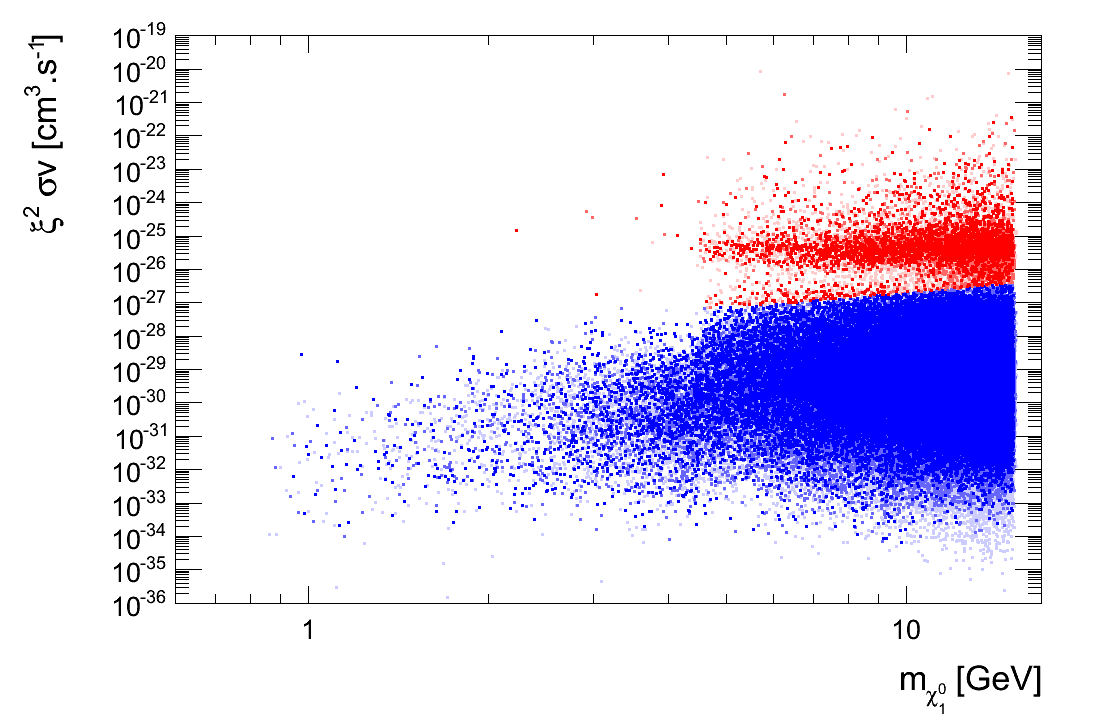}
	\caption{Rescaled neutralino annihilation cross section in the galaxy versus the neutralino mass,  the points 
	which overpredict the gamma ray flux in dSph are in red, see section III. }
\label{fig:sigmaV_v0}
\end{figure}

The total annihilation cross section spans several orders of magnitude as displayed in Fig.~\ref{fig:sigmaV_v0}. 
In some cases, it can be strongly enhanced with respect to its value in the Primordial Universe.
This ''boost'' can occur when the annihilation
proceeds through a s-channel exchange of a pseudoscalar Higgs particle near resonance, the cross section is then sensitive to the thermal kinetic energy : 
at small velocities, one gets the full resonance enhancement while at $v \sim c$, one only catches 
the tail of the resonance~\cite{Bi:2009uj,Ibe:2008ye}.

The associated cross section is proportionnal to 
\begin{eqnarray}
v\sigma(v) &\propto& \frac{1}{(s-m_A^2)^2 +\Gamma_A^2 m_A^2}\nonumber\\
&=& \frac{1}{16m_\chi^4}\frac{1}{(v^2/4+\Delta)^2+\Gamma_A^2
(1-\Delta)/4m_\chi^2}
\label{eq:Delta}
\end{eqnarray}
where $\Delta=1-m_A^2/4m_\chi^2$.  At $v\rightarrow 0$, it is strongly enhanced 
as compared to its value at freeze-out  for  $\Delta,\Gamma_A\ll 1$.  

\begin{figure}[h]
	\centering
\includegraphics[width=0.42\textwidth]{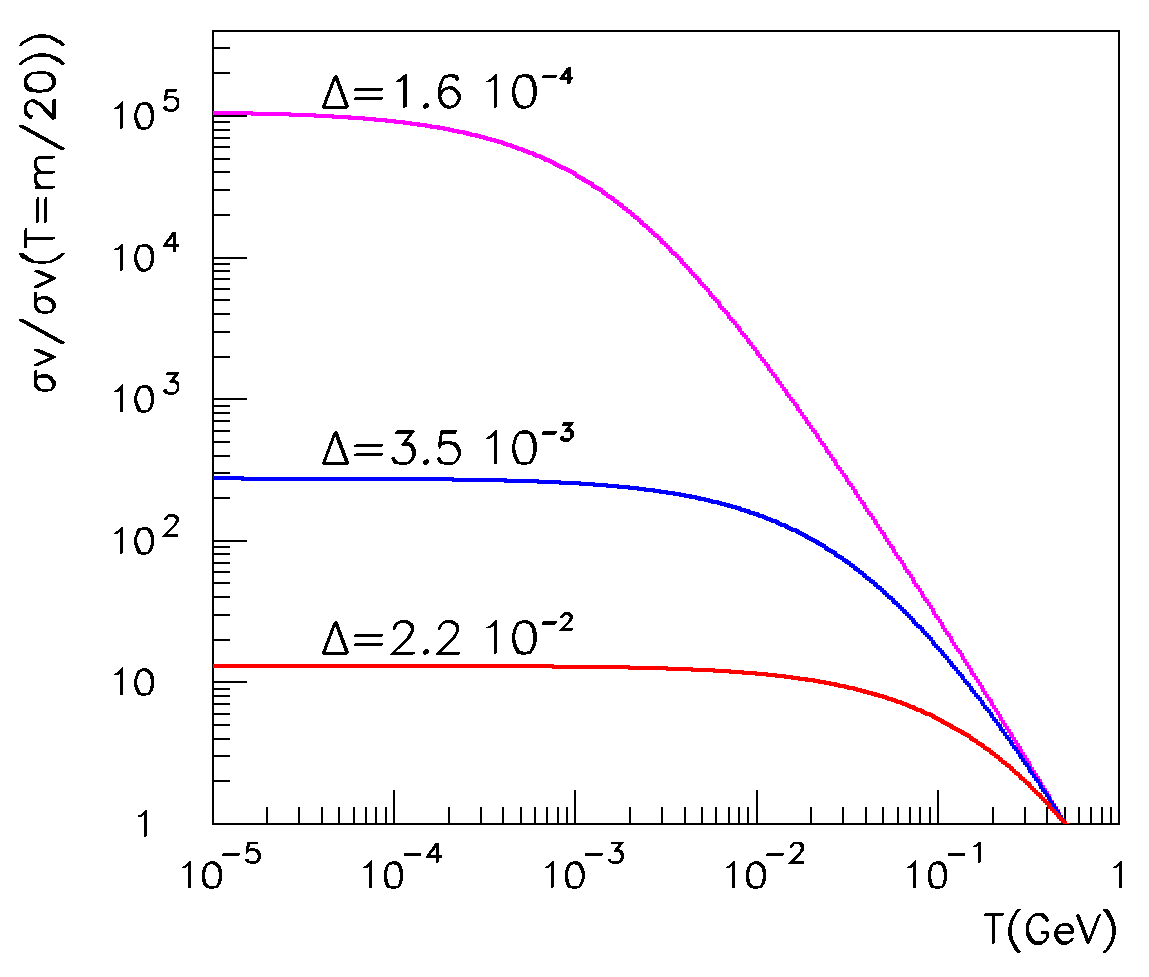}
	\caption{
	Ratio of $\langle \sigma v \rangle$ to $\langle \sigma v\rangle (T=m/20)$ as a function of the temperature.}
\label{fig:boost}
\end{figure}

To give a more quantitative estimate of this effect, 
let us consider one allowed scenario with $m_{\chi}=10.08$~GeV, $m_{A_1}=20.12$~GeV, $\Gamma=1.1\times 10^{-4}$~GeV 
and compute the ratio of the thermally averaged cross section at a given temperature, $\langle \sigma v \rangle (T)$ to the value at a typical 
freeze-out temperature, $\langle \sigma v \rangle (T=m/20)$.  The enhancement factor reaches two orders of magnitude and depends mostly on $\Delta$ since the term in $\Gamma_A$ in Eq.~\ref{eq:Delta} is negligible.
A small variation in $\Delta$ can lead to an even larger enhancement factor, see Fig.~\ref{fig:boost} where the enhancement factor
is displayed for different values of $\Delta$.
 
\begin{figure}[h]
	\centering
\includegraphics[width=0.5\textwidth]{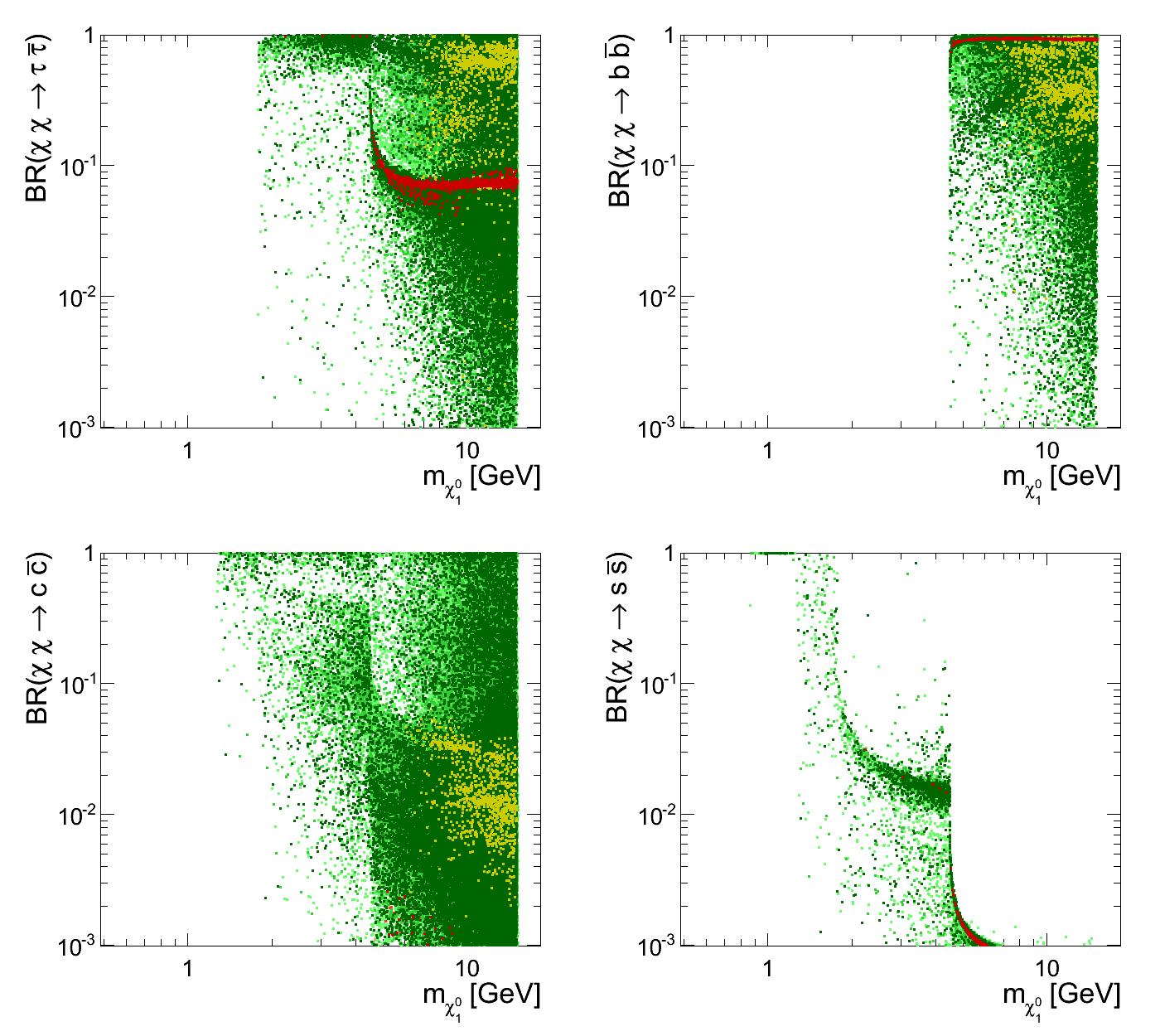}
	\caption{Branching ratios associated with neutralino pair annihilations at low velocities. }
\label{fig:BR_dSPH}
\end{figure}

As a result, we find that the neutralino annihilation branching ratios in the MW or dSph differ
 from those found in the early Universe. 
Our results are summarized in Fig.~\ref{fig:BR_dSPH}, where we only display the final states which opened up at low velocities. 
The $A_1 A_1$ and $H_1 H_1$ are no longer possible final states because the associated cross sections are both suppressed.
On the other hand, the branching ratios for the $c \bar{c}$ and $\tau \bar{\tau}$  now reach unity for many points 
where $m_{\chi} >  m_c$ and $m_{\chi} > m_{\tau}$ respectively.

Many scenarios have indeed a very small annihilation cross section in dwarf galaxies due to a p-wave suppression factor. This affects, in particular, 
the annihilation processes which were dominated in the Early Universe by a $H_1$ resonance decaying into Fermion pairs as well as annihilation into light Higgs final states through the $t$ and $u$ channel neutralino exchange. For these configurations, 
channels such as the t-channel sFermion exchange, which were subdominant in the Early Universe, become important when $v \rightarrow 0$.

In principle, we can make generic predictions for all the configurations with a BR $=1$ for a given channel since  
the annihilation cross section can be factorized. However, most of the dark matter scenarios that we have found 
have a mixture of final states. Hence, in the following, we will compute individually the gamma and cosmic ray fluxes for each point considered.

\section{Astrophysical limits }
We can now investigate the astrophysical limits which apply to
light neutralinos. 

\subsection{Gamma rays in dSph}
\label{sec:dsph}
Objects such as dSph are meant to be dark matter dominated and represent therefore in principle a good target for dark matter 
searches \cite{Mateo:1998wg,Strigari:2006rd,Hooper:2003sh}.
Dark matter annihilations are expected to produce quarks and/or taus in 
the final state which, in turn after hadronization, produce  gamma rays. Photons can also be produced directly as radiation from
an internal line or from a final state. Finally it is also possible that neutralino annihilate directly into photon pairs through a loop-induced process,
but the cross sections are typically small ~\cite{Guillaume}.

The flux of gamma rays originating from dark matter annihilation into all SM final states,  is thus given by
\begin{eqnarray}
\frac{d\phi_{\gamma}(E)}{dE}& =& \frac{1}{2} \frac{1}{4 \pi} \ \left(\frac{(\sigma v)_{tot} }{m_{\chi_0}}\right)^2  \ 
\sum_i BR_i \ \frac{d N_i}{d E} \nonumber\\
&& \xi^2 \int dl(\psi) \  \rho^2_{\rm DM}(l(\psi)) 
\label{eq:flux}
\end{eqnarray}
where $(\sigma v)_{tot}$ is the total annihilation cross section and $BR_i$ is the fraction into a given SM final state $i$,
$\rho_{\rm DM}$ is the dark matter energy density,  
$d N_i/d E$ is the number of photons produced after hadronization 
(or radiation even though this process is generally subdominant) in terms of the energy $E$, 
$l(\psi)$ is the line of sight in the $\psi$ direction. The factor $1/2$  accounts for 
the Majorana nature of the neutralino.

This flux has to be integrated over the Fermi angular resolution. For an angular region of diameter $0.5^{\circ}$, 
as considered by the Fermi experiment \cite{Abdo:2010ex}, one is not sensitive to the inner slope of the dark matter halo. 
Hence we can safely consider a NFW dark matter halo profile. 
The value of the integral along the line of sight and averaged over the resolution is taken from Table 4 in \cite{Abdo:2010ex}.
The energy dependent part of the differential flux is then computed with micrOMEGAs and integrated from $0.1~{\rm GeV}<E<\mlsp$. 

\begin{figure}[h]
	\centering
\includegraphics[width=0.5\textwidth]{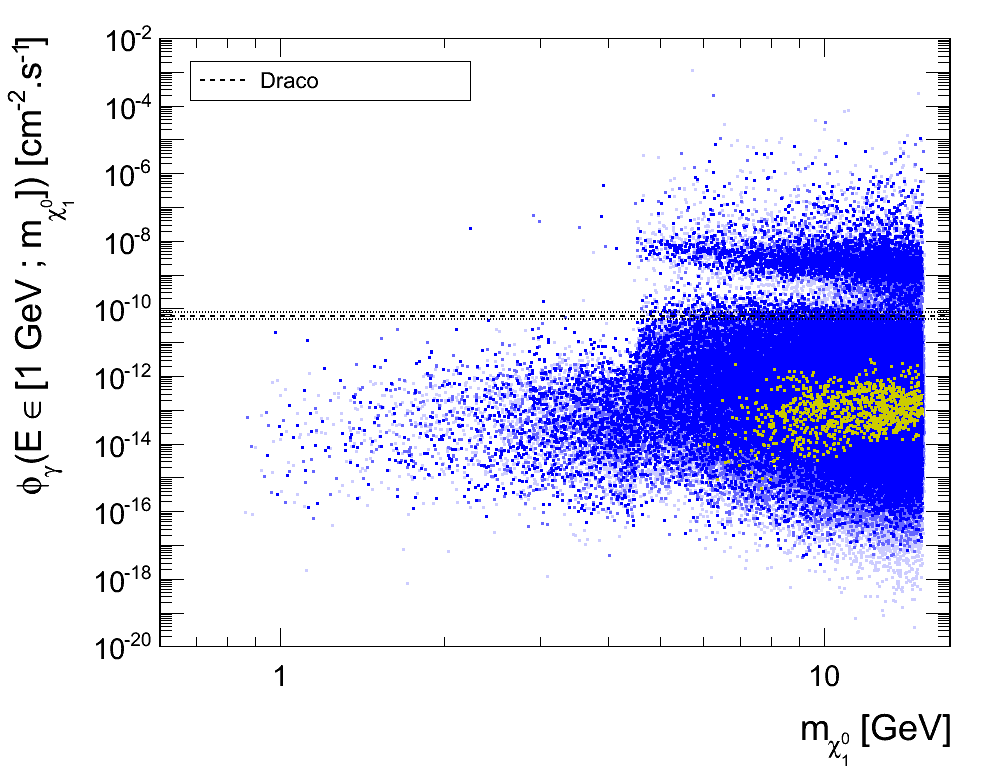}
	\caption{Predicted photon flux as a function of the LSP mass. The horizontal lines correspond to the Fermi limit including the $1\sigma$ error bars in the 
	integral over the DM density distribution.
	Points excluded by XENON100 are in yellow. }
\label{fig:sigmav_vs_mchi_dwarf}
\end{figure}

For each point found by our MCMC, we have computed the gamma ray flux expected in the eight 
dwarfs considered by the Fermi experiment. We then compare this value with the Fermi-LAT 95\% limits ~\cite{Abdo:2010ex}.
Our results are displayed in Fig.~\ref{fig:sigmav_vs_mchi_dwarf} for Draco which gives the most sensitive limits.
As expected, excluded models are those with the largest annihilation cross section (irrespective of the
dominant annihilation channel, see also Fig.~\ref{fig:sigmaV_v0}) and for which the mass difference 
between the LSP and the pseudoscalar involves a fair amount of fine tuning.

Note that the exclusion limit varies only slightly with the mass. Basically, it 
corresponds to $\xi^2\sigma v/m_\chi^2 \ge  2-5 \times 10^{-29} {\rm cm}^3 {\rm s}^{-1} {\rm GeV}^{-2}$.  
Also the criterion for exclusion is based on a comparison of the computed flux with the Draco limit taking into account the $1\sigma$ error bars in the integral over the DM density distribution. Thus, it is a conservative exclusion criterion. However, many points excluded by DRACO are also excluded by other astrophysical observations (including by the flux of antiprotons measured at Earth position). The implications for the Higgs spectrum will be discussed in subsection \ref{sec:higgs}.

\subsection{Antiproton and positron fluxes in the Milky Way}

We can now investigate the cosmic ray production in the MW.
Calculations are similar to gamma rays except that now we focus on positron and antiproton production.
We also expect that the best signals are associated with the largest annihilation cross section. 

Since the limits from dSphs are already very constraining, we shall consider other astrophysical bounds 
as a complementary tool and thus apply them only to the scenarios which survived the Fermi constraints.
For this purpose, we group all the scenarios in  bins of 1 GeV for neutralino masses ranging from 1 to 15 GeV
and, for each of the 14 bins, we select the point with the maximum value for $\xi^2 \sigma v/\mlsp^2$ and with a good likelihood (defined as $Q>0.32 Q_{max}$) and which, yet, is safe with respect to the Draco limits. 
This leads to 14 benchmark points. The dominant annihilation channel for all of them is into $b\bar{b}$ for $\mlsp> 4.2{\rm GeV}$ and into $\tau\bar{\tau}$ otherwise. The benchmarks are listed in Table~\ref{tab:Benchmark}.

\begin{figure}[h]
	\centering
		\includegraphics[width=0.47\textwidth]{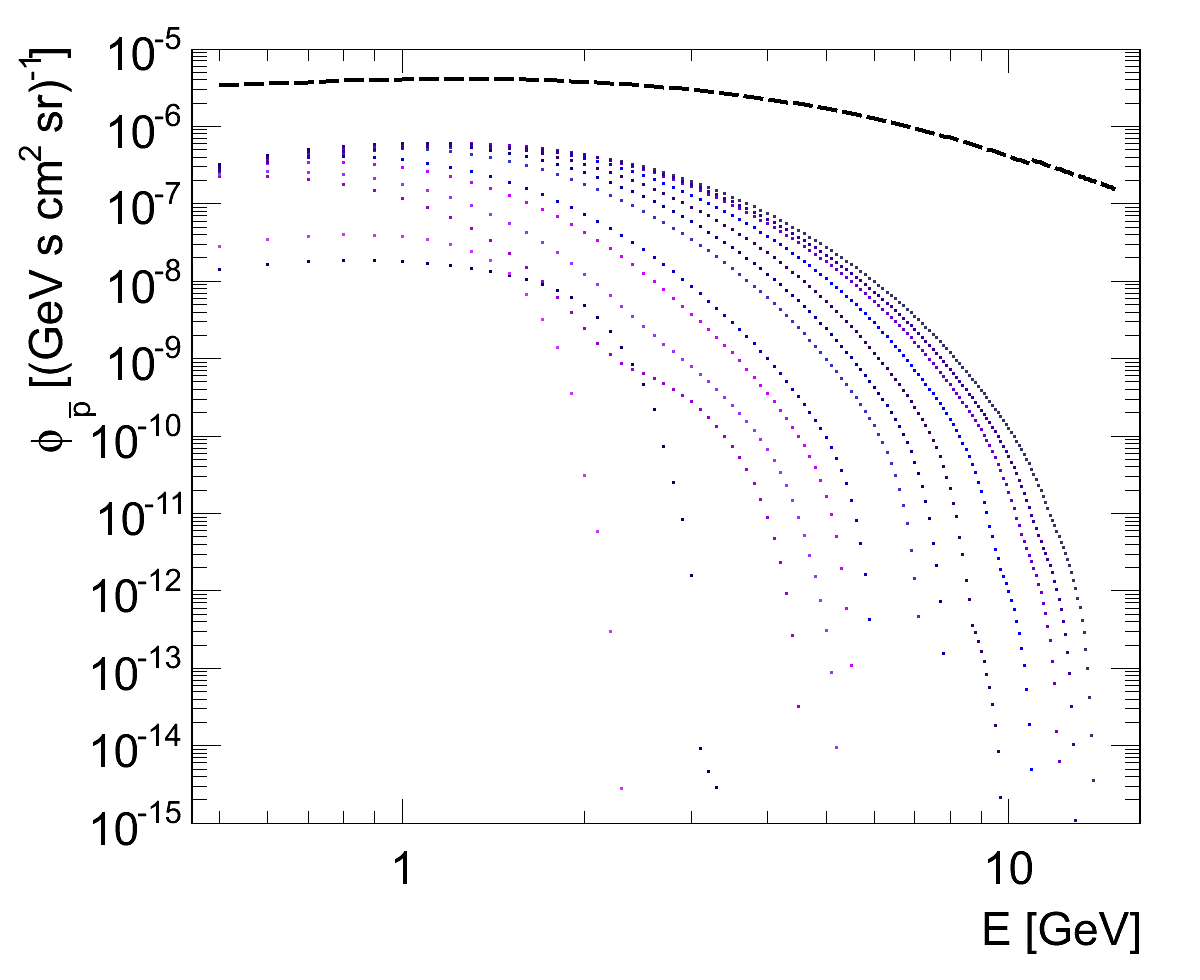}
\caption{Cosmic ray predictions for the antiproton spectra for 13 benchmark points in Table~\ref{tab:Benchmark}. The first point
in this table is ignored as it is too light to give antiprotons. 
The maximal energy for each spectrum corresponds to the neutralino mass.  The background from 
secondaries~\cite{Maurin:2006hy} (dash) is also displayed. 
}
\label{fig:CR_vs_Mchi_Dwarf}
\end{figure}

The results for the antiproton fluxes are displayed in Fig.~\ref{fig:CR_vs_Mchi_Dwarf} and compared to the
background spectrum taken from the approximate analytical formulae in Ref~\cite{Maurin:2006hy}. This background by itself provides a good
fit to the data. 
The fluxes computed are close to one order of magnitude below the background from secondaries at energies below $\approx 2$~GeV 
and drop rapidly at higher energies. 
We therefore conclude that one cannot further constrain  these scenarios by measuring the antiproton flux as
the uncertainty in the background calculation (which exceeds 10\%) is always larger than the signal. 

Note that to compute these fluxes we have set the propagation parameters to the default values in micrOMEGAs, i.e. the
MED set of parameters (see Ref.~\cite{Donato:2003xg}). 
For a different set of propagation parameters, the expected fluxes can increase. Indeed we find that the fluxes reach at most
the background  level with the MAX set of propagation parameters (corresponding to a larger diffusive zone, see Ref.~\cite{Donato:2003xg,Delahaye:2007fr}). 

\begin{figure}[h]
	\centering
\includegraphics[width=0.47\textwidth]{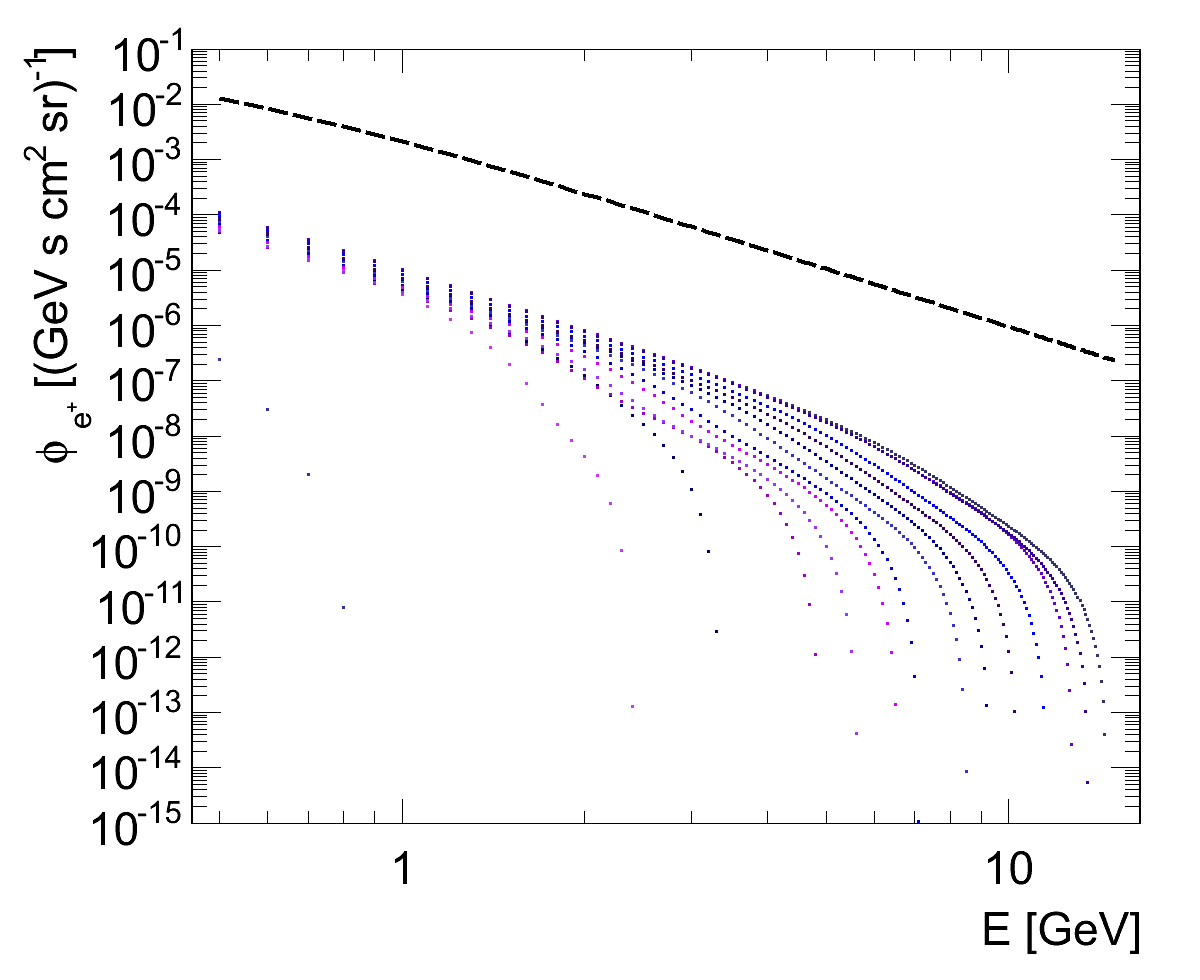}
\caption{ Predictions for the  positron spectra as compared to the background~\cite{Baltz:1998xv} (dash) for 14 benchmark points in Table~\ref{tab:Benchmark}.
The maximal energy for each spectrum corresponds to the neutralino mass.
}
	\label{fig:pos_vs_Mchi_Dwarf}
\end{figure}

For completeness, we have also computed the positron fluxes for the 14 benchmark points, using the MED propagation parameter set \cite{Delahaye:2007fr}.  The fluxes are always at
least two orders of magnitude below the background~\cite{Baltz:1998xv} , see Fig.~\ref{fig:pos_vs_Mchi_Dwarf}.
This was to be expected since the scenarios that we have selected predict dominant quark final states; 
hence a better signature in antiprotons than in positrons. Note that the $\tau\bar{\tau}$ channel, which leads to a hard positron
spectrum, is dominant either at very low masses or when the cross section is very small. 

\begin{table}
\begin{center}
\begin{tabular}{|c|c|c|c|c|c|c|c|c|}
\hline
$\rm{m_{\chi_1^0}}$ &  $\rm{\xi}$ & $\langle \sigma v \rangle \times 10^{27}$ & $\rm{BR_{\tau \bar{\tau}}}$ & $\rm{BR_{b \bar{b}}}$ & 
$\rm{BR_{s \bar{s}}}$ & R \\
$\rm{[GeV]}$ &   & $\rm{[cm^3s^{-1}]}$ &  &  & & \\ 
\hline
\hline
 0.976  & 0.373 & 0.209& 0 & 0  & 0.997 & 0   \\
\hline
 2.409  & 1.00      & 0.297  & 0.964 & 0  & 0.026 & 0.040 \\
\hline
 3.342  & 0.935 & 0.345  & 0.972 & 0  & 0.018  &0.044 \\
\hline
 4.885  & 0.465 & 3.298  & 0.0970 & 0.901  & 0.0016 &0.041 \\ 
\hline
 5.626 & 0.376 & 5.389  & 0.0698 & 0.929 & 0.0011 &0.040 \\
\hline
 6.551 & 0.528 & 3.547  & 0.0618 & 0.937  & 0 &0.046 \\
\hline
 7.101  & 0.689 & 2.425  & 0.0586 & 0.940  & 0 &0.050 \\
\hline
 8.513  & 0.829 & 2.161  & 0.0416 & 0.958  & 0 &0.055 \\
 \hline
 9.274  & 0.827 & 2.497  & 0.0533 & 0.946  & 0 &0.060 \\
\hline
 10.27  & 0.906 & 2.323  & 0.0634 & 0.935  & 0 &0.063 \\
\hline
 11.50 & 0.960 & 2.575  & 0.0611 & 0.937  & 0 &0.074 \\
\hline
 12.74 & 0.955 & 3.224 & 0.102 & 0.897 & 0&0.088 \\
\hline
 13.51 & 0.558 & 9.571  & 0.0781 & 0.921 & 0 &0.085 \\
\hline
 14.48  & 0.147 & 148.4  & 0.0748 & 0.924  & 0 &0.088 \\
\hline
\hline
\end{tabular}
\end{center}
\caption{Benchmark points: main characteristics and ratio of the dark radio emissivity at 330MHz to observation (R).}
\label{tab:Benchmark}
\end{table}

\subsection{Comparison with  direct detection}

\begin{figure}[h]
	\centering
		\includegraphics[width=0.5\textwidth]{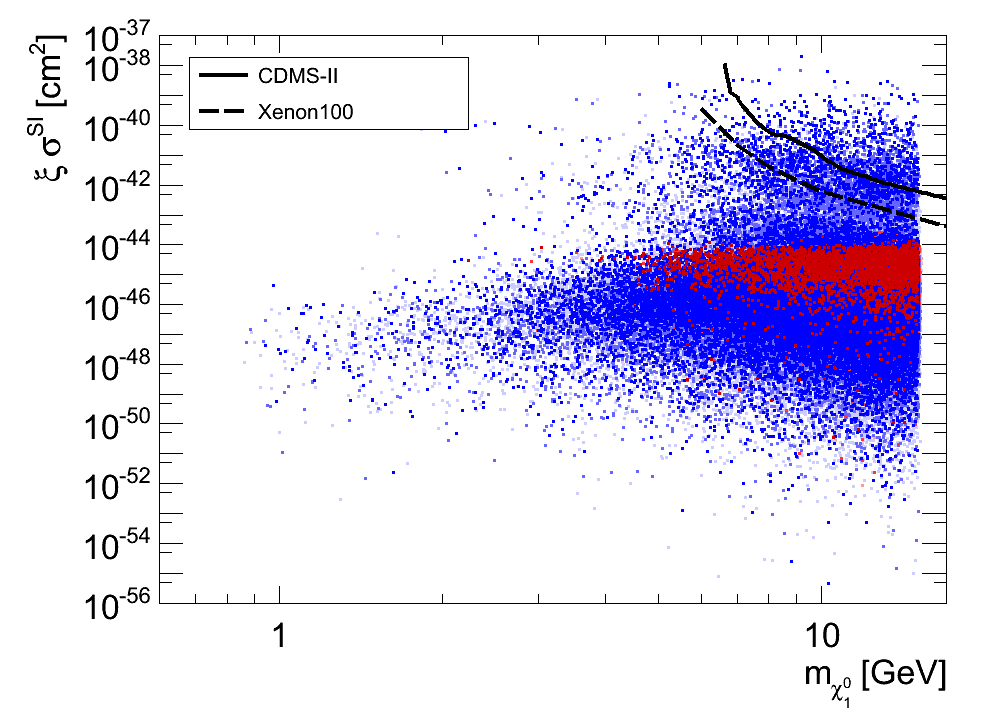}
\caption{Spin independent cross section versus the neutralino mass. 
In red are the points which over predict the gamma ray flux in dSph.}
\label{fig:SI_vs_Mchi_Dwarf}
\end{figure}

\begin{figure}[h]
	\centering
		\includegraphics[width=0.5\textwidth]{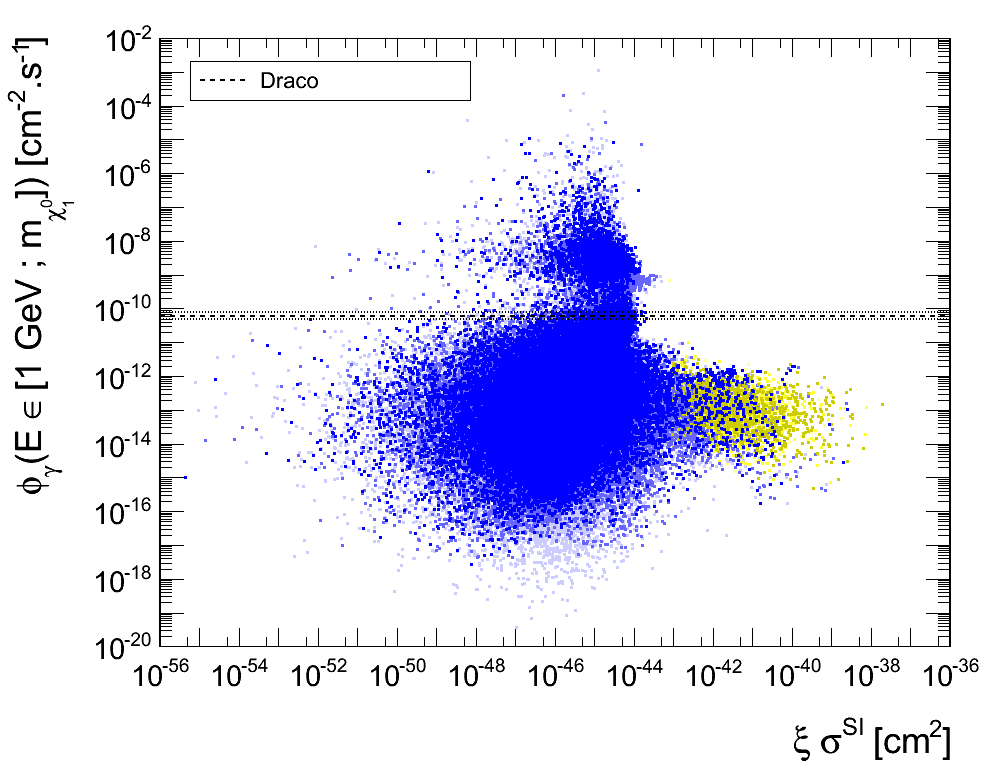}\\
\includegraphics[width=0.5\textwidth]{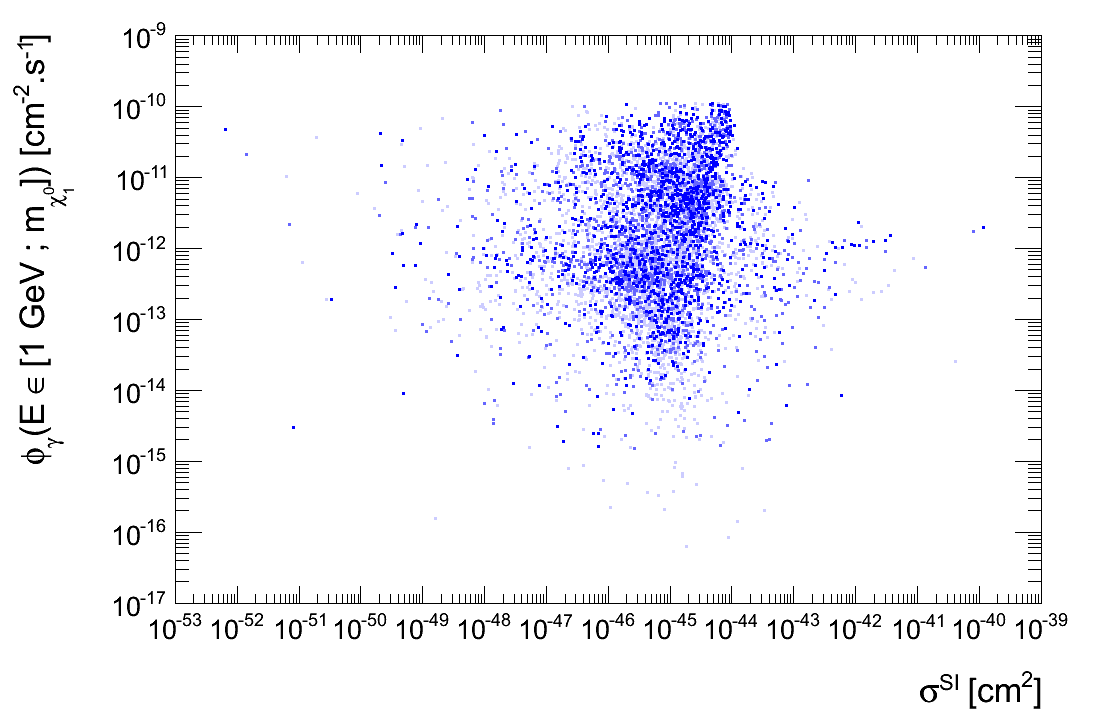}
\caption{Correlation between the gamma ray flux and spin independent cross section of NMSSM neutralinos. 
Top: all points are included. The yellow points correspond to scenarios with a too large spin independent 
cross section. The dashed line corresponds to the Fermi limit for the Draco dSph. 
Bottom: all points overpredicting the gamma ray flux or 
with a too large spin independent cross section and which do not completely explain  the dark matter today have been removed.} 
	\label{fig:Excludedpoints}
\end{figure}

In the light neutralino scenarios there is a good complementarity between gamma ray searches and direct searches. 
Indeed many scenarios which  predict a spin independent cross section below the XENON100 exclusion curve 
overpredict the gamma ray flux in dSph galaxies, see Fig.~\ref{fig:SI_vs_Mchi_Dwarf}. 
This is also illustrated in the correlation plot displayed in upper panel of Fig.~\ref{fig:Excludedpoints}, 
where we show the gamma ray flux as a function of the spin independent cross section.
Clearly, the Fermi dSph limits constrain scenarios where  
the spin-independent cross section is smaller than the latest XENON100 limit while the 
XENON100 limits exclude points where the gamma ray flux in dSph is not yet accessible by the Fermi searches. 
In the framework of the NMSSM, this complementarity is directly connected to the light Higgs spectrum as discussed in the next subsection.

If we now remove the points which do not have the correct abundance today and exclude the points which produce 
too many gamma rays in Draco and non observed events in XENON100 (see Fig.\ref{fig:Excludedpoints}, lower panel), 
we obtain that, statistically, 
light neutralinos are likely to produce 
$[10^{-14},10^{-10}] \ \rm{\gamma/cm^2/s}$ and have a spin independent cross section of 
$[10^{-48},10^{-44}] \ \rm{cm^2}$.

We have also computed the gamma ray flux for points which are in the region favoured by the CoGeNT experiment ~\cite{Aalseth:2010vx}. All these points lie in the region excluded by Xenon100. However, since CoGeNT claims 
detection at 2 $\sigma$ of an annual modulation signal \cite{Aalseth:2011wp}, it is worth investigating the astrophysical limits for such candidates.

Since we have demonstrated that indirect and direct detection experiments were probing different regions of the parameter space and these candidates are within XENON100 sensitivity, we do not expect that they produce large gamma ray and cosmic ray fluxes. However, to check this statement, we shall consider three benchmark points (cf Table \ref{tab:NMSSM_M15_CoGenT}).

\begin{table}[htp]
\begin{center}
\begin{tabular}{|c|c|c|c|c|c|c|c|c|c|c|}
\hline
$\rm{M_1}$ & $\rm{M_2}$ & $\rm{M_{\tilde{l}}}$ & $\rm{M_{\tilde{q}}}$ & $\rm{\mu}$ & $\rm{tan\beta}$ & $\rm{\lambda}$ & $\rm{\kappa}$ & $\rm{A_{\lambda}}$ & $\rm{A_{\kappa}}$ & $\rm{A_t}$ \\
\hline
14.6 & 1257 & 166 & 1284 & 175 & 20.7 & 0.55 & 0.27 & 3529 & -361 & 1005 \\
\hline
22.3 & 157 & 528 & 1701 & 164 & 20.0 & 0.55 & 0.15 & 3281 & -212 & 1591 \\
\hline
16.8 & 605 & 192 & 1782 & 186 & 18.3 & 0.70 & 0.25 & 3464 & -317 & 2437 \\
\hline
\end{tabular}
\end{center}
\caption{Three examples of NMSSM points falling in the CoGenT contour~\cite{Aalseth:2010vx} in the $\xi \sigma^{SI}$ vs. $m_{\chi_1^0}$ plane. For all of them we have set $M_3=3\times M_2$ and $A_b=A_{\tau}=0$. All quantities are expressed in $GeV$ units.}
\label{tab:NMSSM_M15_CoGenT}
\end{table}

For these points, we found $\xi^2 \ \sigma v/m_\chi^2 \ \le \ 6 \times 10^{-31} {\rm cm}^3 {\rm s}^{-1} {\rm GeV}^{-2}$ which is one or two orders of magnitude below the Draco limit in section ~\ref{sec:dsph}. Hence, it seems that NMSSM neutralinos in the CoGeNT region are not excluded by the indirect detection limits for the moment.

This is an important point nevertheless. Indeed, it will be challenging for both dark matter direct and indirect detection experiments to reach the level of sensitivity which is required to completely probe the region of the parameter space with light neutralinos. 
In addition, new sources of background (such as the neutrino background 
for the direct detection experiments) may also weaken the analysis.

\subsection{Implication for particle physics}
\label{sec:higgs}

In the previous subsections, we have demonstrated that the Fermi dSph limits were setting stringent limits on the 
NMSSM parameter space and were complementary to dark matter direct detection searches.  
We can now examine the impact of these limits on the Higgs sector and on B-physics observables. 

\begin{figure}[h]
	\centering
\includegraphics[width=0.45\textwidth]{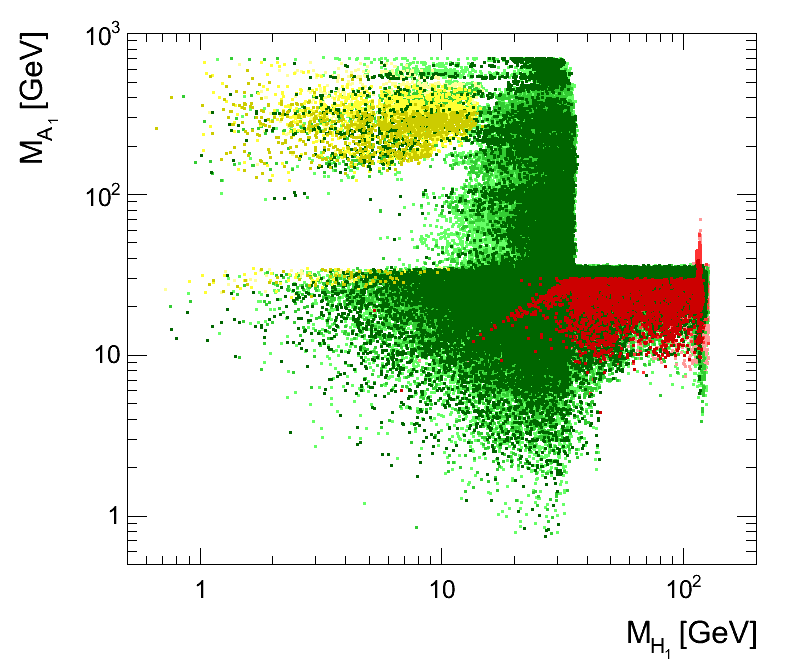}
	\caption{Correlation in the ($m_{A_1}$,$m_{H_1}$) plane. All the safe configurations
	(green) as well as   the excluded points by XENON100 (yellow) and dSph (red) are displayed.}
\label{fig:H1A1}
\end{figure}

\begin{figure}[h]
	\centering
\includegraphics[width=0.5\textwidth]{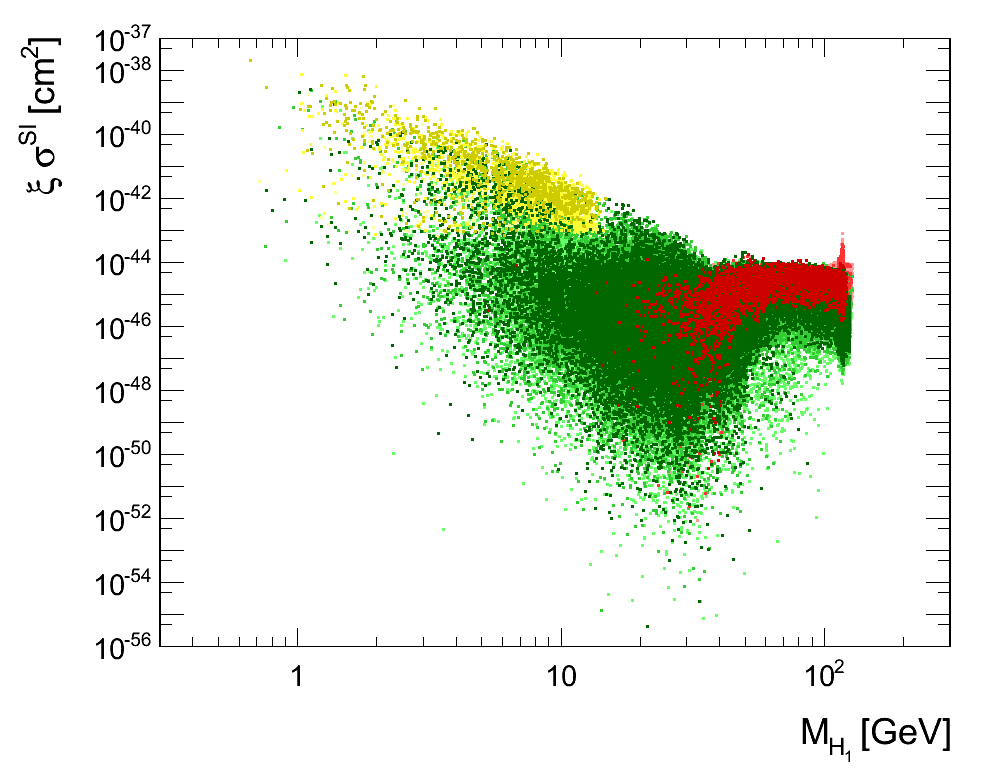}
	\caption{Spin independent cross section in terms of $m_{H_1}$, same color code as Fig.~\ref{fig:H1A1}.} 
\label{fig:SIcs_vs_Higgses}
\end{figure}

Efficient neutralino annihilation in the Early Universe requires  at least one light Higgs ($m_{H_1},m_{A_1}<30$~GeV for
$m_{\tilde\chi}<15$~GeV) as illustrated in Fig.~\ref{fig:H1A1}. Astrophysical limits then apply in two distinct regions
of the $m_{A_1}-m_{H_1}$ plane. 
The first region corresponds to a light $H_1$ ($m_{H_1} \in [1,10] $ GeV) and to heavier
$A_1$ (with $m_{A_1} \in [10,1000]$ GeV). In this region 
the spin-independent cross section 
can become very large, which is in conflict with XENON100 data (in yellow in Fig.~\ref{fig:H1A1}). 
Indeed, as can be seen in Fig.~\ref{fig:SIcs_vs_Higgses}, 
 larger spin independent cross section are found for light $H_1$. This is because the scalar exchange contribution to the cross section
goes as $1/m_{H_1}^4$ . Note that because sufficiently large couplings of the light Higgs
to the LSP and to quarks in the nucleon are necessary to have a large SI cross section~\cite{Vasquez:2010ru}, 
 many points with light $H_1$ are not excluded.
The second region 
corresponds to a relatively light $A_1$ ($m_{A_1} \in [10,30]$ GeV) and 
$m_{H_1} \in [20,100]$ GeV. Here
the neutralino pair annihilations  (which proceed through the exchange of an $A_1$ in the  s-channel) 
become singular when $v_{dm} \rightarrow 0$ and $1-m_{A_1}/2m_{\chi}<<1$ and thus can produce 
too many gamma rays in dwarf galaxies (see points in red in Fig.~\ref{fig:SIcs_vs_Higgses}).

\begin{figure}[h]
	\centering
\includegraphics[width=0.5\textwidth]{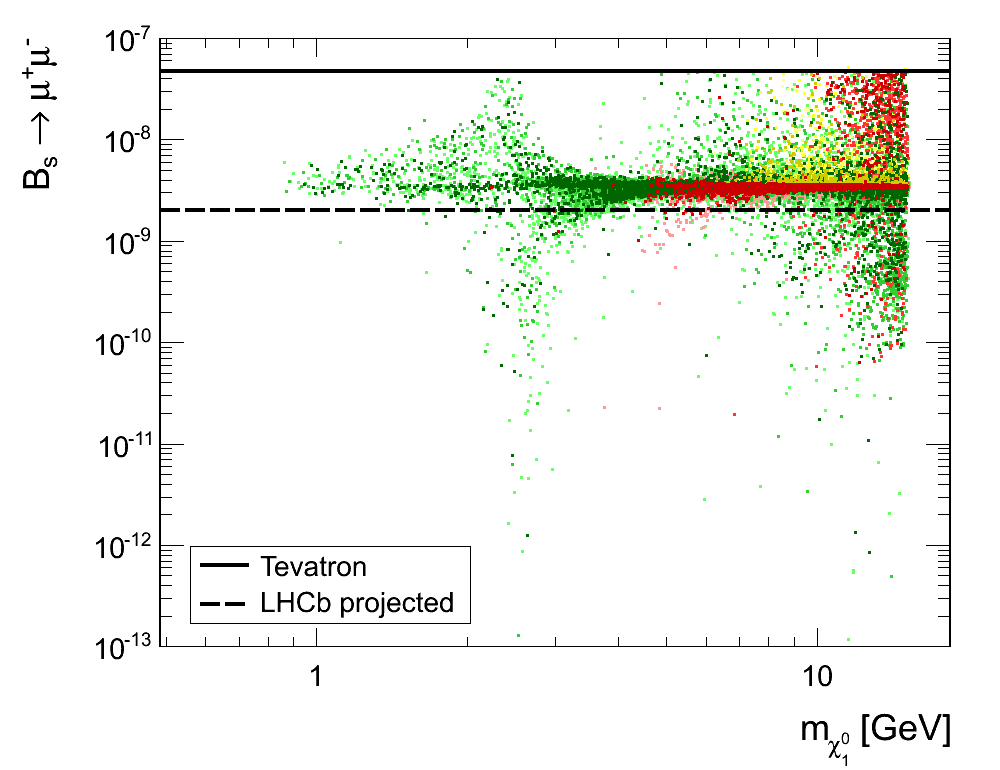}
	\caption{Predictions for the branching ratio $B(B_s\rightarrow \mu^+\mu^-)$ in terms of $m_{\chi_1}$ together with the Tevatron limit and the LHCb projected limit, same color code as Fig.~\ref{fig:H1A1}.} 
\label{fig:Bsmumu_vs_mchi}
\end{figure}

The observables in the B-sector have been used to constrain the parameter space. The LHCb experiment, which is now taking data, will measure these observables with increased precision, it is therefore interesting to examine whether this will probe further our scenarios. For example the branching ratio for $B_s\rightarrow \mu^+\mu^-$ is expected to be  powerful in probing scenarios with a light doublet Higgs and large values of $\tan\beta$. Only a fraction of our scenarios have such characteristics.
The predictions for $B(B_s\rightarrow \mu^+\mu^-)$ are displayed in Fig.~\ref{fig:Bsmumu_vs_mchi} together with the expected limit of LHCb with ${\cal L}=1fb^{-1}$~\cite{lhcb}. A signal is expected only for a small fraction of the scenarios while in some cases the predictions are suppressed as compared to the SM expectation ($B(B_s\rightarrow \mu^+\mu^-)= 3.6\pm 0.4\times 10^{-9}$)~\cite{Buras:2009if}.  This implies that light neutralino scenarios cannot be probed entirely with this observable.

\section{Radio emission in the galactic centre and the Coma cluster}

Since light particles eventually produce electrons in the MW, one expects a significant radio emission in 
the galactic centre as well as in galaxy clusters as first pointed out in \cite{Boehm:2002yz} and 
discussed further in e.g. Ref.~\cite{Colafrancesco:2005ji, Borriello:2008gy,Crocker:2010gy,Colafrancesco:2010kx,Boehm:2010kg}. 

This is true, in particular, for the points with large cross sections at $v \rightarrow 0$ and $\xi \simeq 1$ (i.e. canonical values in the primordial Universe). 
Since these points  have been already excluded by using the Fermi dSph limits, we shall rather concentrate on the part of the parameter space that is left after 
applying the gamma ray constraint.  We thus compute the radio flux expected for the benchmark points displayed in Table \ref{tab:Benchmark}. 

We repeat the same calculations as in Ref.~\cite{Boehm:2010kg} where we assume the MED set of propagation parameters 
and set the magnetic field, $B$, to 20 $\mu$G. This value is slightly higher than that derived in ~\cite{LaRosa:2005ai} for 
very small scales but it is still conservative enough with respect to the very large value generally considered for the magnetic field 
in the galactic centre. Choosing the MAX set of propagation parameter 
would in fact decrease the intensity of the radio emission expected in the galactic centre but it would also 
lead to a broader radio emission in the galaxy. On the contrary, for the MIN set of propagation parameters, 
one expects a brighter emission in the galactic centre which should be easier to constrain in principle.

We focus on one radio frequency, namely 330 MHz where the emission is about 360 Jy (for an angular resolution of 7') and compute the ratio $R$ of the ``dark'' emissivity to the observation at this frequency in the galactic centre, for each of the points mentioned above. The results are displayed in Table \ref{tab:Benchmark}, in the last column. We find that all the benchmark points have a  negligible radio flux.  This can be explained by the fact that, in all these scenarios, 
the total number of electron produced in the 0.5 to 15 GeV range is about unity  (a minimal injection energy of 1 GeV is required for $B=20 \ \mu$G to produce synchrotron emission at 330 MHz).  Indeed, all the benchmark points annihilate preferably into b or $\tau$-pairs. This leads to a very small number of electrons per energy unit for $E > 1$ GeV.

Note that the determination of the magnetic field value is of particular importance because it changes the synchrotron emission (hence the radio predictions) and the losses. However, it also determines other possible signatures such as an anomalous submillimetre emission in the galaxy which could contribute as a new source of foreground in Cosmological Microwave Background studies. For example, to get a signal at 33 GHz (relevant for both WMAP and Planck) from a 10 GeV dark matter particle, the magnetic field should be greater than $25 \mu$G~\cite{Delahaye:2011jn}. 
If such a value is indeed attained in the galactic centre or in the outer parts, one can hope to 
correlate "dark" radio emission with WMAP and Planck observations e.g. \cite{Hooper:2007kb,McQuinn:2010ju,Bottino:2008sv,Delahaye:2011jn}.

One can also estimate the radio emission in clusters of galaxies to determine whether light dark matter particles 
are potentially constrained by radio observations in these objects \cite{Boehm:2002yz,Boehm:2010kg,Colafrancesco:2005ji}.  
To compute the flux, we consider the Coma cluster and extend the procedure described in \cite{Boehm:2002yz,Boehm:2010kg} to account for the energy distribution of electrons. We assume a NFW profile~\cite{Navarro:1996gj} with $\rho_0 =  4.4 \ 10^{-2} \ \rm{GeV/cm^{3}}$, $r_s = 400$ kpc, a detector angular resolution of 1 deg, a magnetic field of 4.7 $\mu G$ and a density of electrons of $3 \ 10^{-3} \ \rm{cm^{-3}}$. 

\begin{figure}
	\centering
		\includegraphics[width=0.5\textwidth]{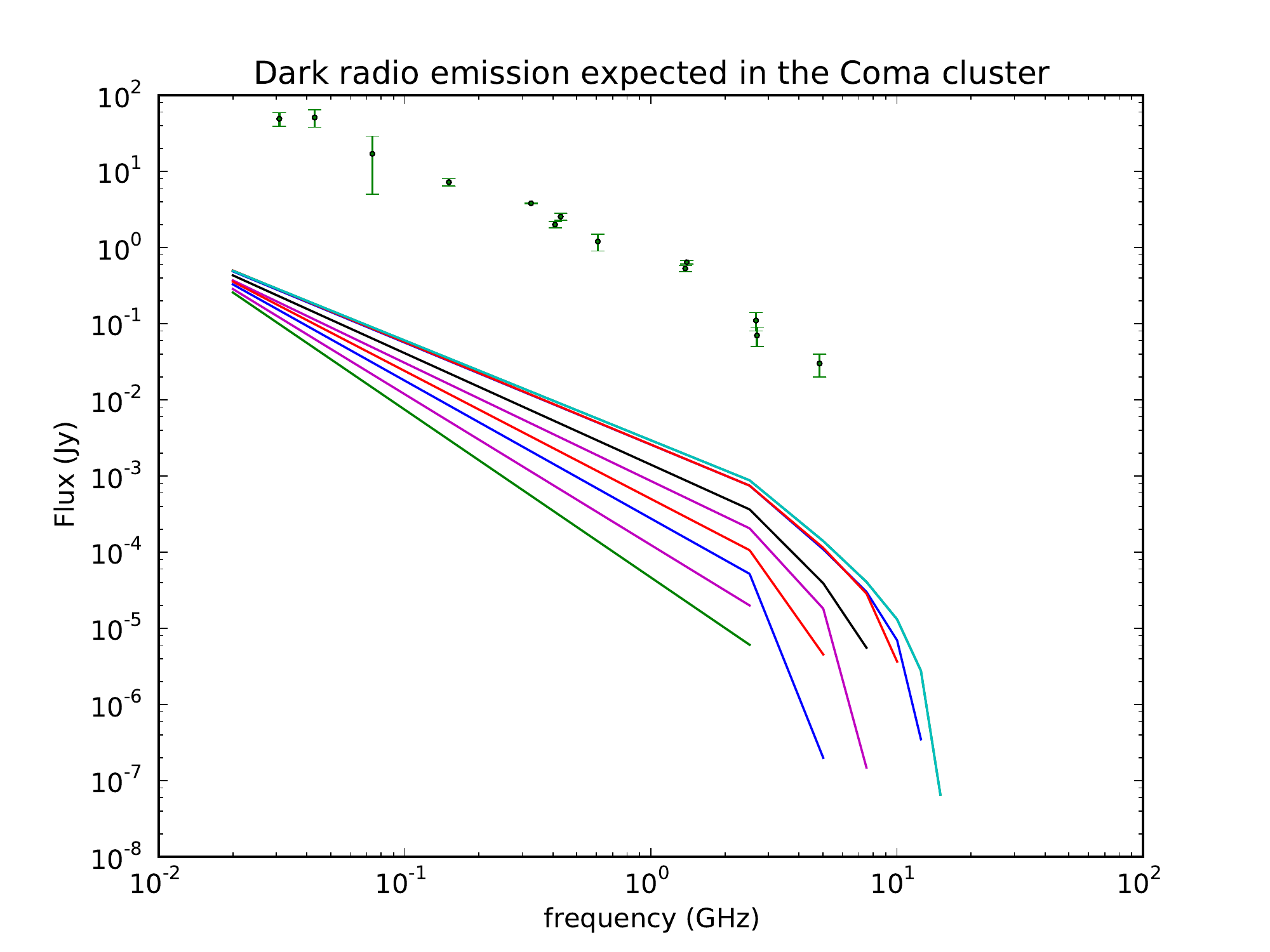}
		\caption{Radio flux in the Coma cluster for 13 benchmark points displayed in Table.~\ref{tab:Benchmark} from the 
		second lightest (left) to the heaviest (right) candidates.  The dataset are taken from the references in \cite{Thierbach:2002rs}.}
	\label{fig:flux_radiocluster_benchmark2}
\end{figure}

The results for our benchmark points are displayed in Fig.\ref{fig:flux_radiocluster_benchmark2} \footnote{ We have checked that we could recover the results in \cite{Colafrancesco:2005ji} for a 40 GeV candidates annihilating into a $b \ \bar{b}$ pair and with the same cross section.}. Our prediction shows that none of these points are excluded by the observation in the Coma cluster, as expected already from our computations in the galaxy.

These results do not account for substructures. Also, they assume a specific value of the magnetic field. If we increase this value by a significant amount, the radio flux becomes larger and the emission becomes also possible at higher frequencies. For example, increasing the value of the magnetic field up to 12 $\mu$ G increases 
the radio flux at 4.58 GHz by a factor $\sim$ 4.5 for the last benchmark point (corresponding to a candidate with  14.48 GeV mass). Still this is not enough to rule out this candidate.

\section{Conclusion}

In this article, we have investigated the astrophysical limits on light NMSSM neutralinos. 
We have shown that the Fermi searches for gamma rays from dwarf Spheroidal galaxies 
set stringent limits on the NMSSM parameter space. In addition, combining these results 
with indirect and direct detection searches restrict 
the parameter space even further. Yet light neutralinos are not ruled out.

It may be extremely difficult to completely probe   light neutralinos 
in the forthcoming future. Indeed, this would require to improve both direct and indirect detection 
experiments sensitivity by several order of magnitudes while experiments may run into new background sources.

The LHC could nevertheless provide crucial information on the existence and on the spectrum of supersymmetric particles and investigate distinctive signatures
involving light neutralinos, for example invisible decay modes of the Higgs~\cite{Cao:2011re}.
Furthermore the mass of a light neutralino, produced in the decay of a selectron, could be determined at a future linear collider, for this selectrons have to be light enough to be pair produced ~\cite{Conley:2010jk}.
Light DM particles can also have an impact on the CMB \cite{Boehm:2002yz,Galli:2009zc,Galli:2011rz,Hutsi:2011vx}. 
In Ref~\cite{Hutsi:2011vx,Galli:2011rz}, it was shown that their annihilation cross section is 
constrained by WMAP to be around the usual 
freeze-out cross section unless DM particles annihilate mainly into neutrinos, as in \cite{Boehm:2006mi} for even lighter DM candidates. 
However, the Planck satellite should further probe these scenarios, including by detecting an anomalous synchrotron emission \cite{Delahaye:2011jn}.
Nonetheless, at present, our results demonstrate that the complementarity between direct and indirect searches 
is a reality and a necessity, at least in the framework of supersymmetric scenarios.

\section{Acknowledgments}
GB thanks the LPSC for its hospitality. The authors would like to thank David Maurin for information on 
the flux  of antiprotons from secondaries, Guillaume Chalons for computing the gamma-ray line for  
several of our scenarios as well as  Alexander Pukhov, Joe Silk and Pierre Salati 
for useful discussions. This work was supported in part by the CNRS-PICS grant ``Propagation of low energy positrons''.

\bibliography{nmssm1}
\end{document}